\documentclass[aps,prd,showpacs,preprintnumbers,nofootinbib,floatfix,floats,groupedaddress,twocolumn]{revtex4-1}

\usepackage{bm}
\usepackage{latexsym}
\usepackage{dcolumn}
\usepackage{amsmath,amsfonts,amssymb}
\usepackage{graphicx,epsfig}
\usepackage{color}
\usepackage[active]{srcltx}
\usepackage{subfig}
\usepackage{slashed}
\usepackage{tikz}
\usetikzlibrary{shapes,snakes}
\usepackage{dsfont}
\usepackage[font=normal,labelfont=bf]{caption}
\def\nn{\nonumber}

\def\l{\left}
\def\r{\right}
\def\DM{\mathrm{d}}

\newcommand{\gae}{\lower 3pt \hbox{$\,\, \buildrel {\scriptstyle >}\over {\scriptstyle
\sim}\,\,$}}
\newcommand{\lae}{\lower 2pt \hbox{$\, \buildrel {\scriptstyle <}\over {\scriptstyle
\sim}\,$}}

\reversemarginpar

\def \lp { \ell_0}

\def \ge {\widehat{\bm g}}

\begin{document}

\title{Action and Observer dependence in Euclidean quantum gravity
}

 \author{Dawood Kothawala}
 \email{dawood@iitm.ac.in}
 \affiliation{Department of Physics, Indian Institute of Technology Madras, Chennai 600 036}

\date{\today}
\begin{abstract}
\noindent
Given a Lorentzian spacetime $\l( \mathcal{M}, \bm g \r)$ and a non-vanishing timelike vector field $\bm u(\lambda)$ with level surfaces $\Sigma$, one can construct on $\mathcal{M}$ a Euclidean metric $\bm g_E^{-1} = \bm g^{-1} + 2 \bm u \otimes \bm u$ [Hawking \& Ellis, 1973]. Motivated by this, we consider a class of metrics $\bm \ge {}^{-1} = \bm g {}^{-1} - \Theta(\lambda)\, \bm u \otimes \bm u$ with an arbitrary function $\Theta$ 
that interpolates between the Euclidean ($\Theta=-2$) and Lorentzian ($\Theta=0$) regimes, separated by the codimension one hypersurface $\Sigma_0$ defined by $\Theta=-1$.
Since $\bm \ge$ can not, in general, be obtained from $\bm g$ by a diffeomorphism, its Euclidean regime is in general different from that obtained from Wick rotation $t \rightarrow - i t$. For example, if $\bm g$ is the $k=0$ Lorentzian de Sitter metric corresponding to $\Lambda>0$, the Euclidean regime of $\bm \ge$ is the $k=0$ Euclidean anti-de Sitter space with $\Lambda<0$. 
\\
\\
\noindent We analyze the curvature tensors associated with $\widehat{\bm g}$ for arbitrary Lorentzian metrics $\bm g$ and timelike geodesic fields $\bm u$, and show that they have interesting and remarkable mathematical structures: 
(i) Additional terms arise in the Euclidean regime $\Theta \to -2$ of $\bm \ge$. 
(ii)  For the simplest choice of a {\it step}-profile for $\Theta$, the Ricci scalar \textbf{Ric}$[\widehat{\bm g}]$ of $\bm \ge$ reduces, in the Lorentzian regime $\Theta \to 0$, to the complete Einstein-Hilbert lagrangian {\it with the correct} Gibbons-Hawking-York boundary term; the latter arises as a {\it delta}-function of strength $2K$ supported on $\Sigma_0$. 
(iii) In the Euclidean regime $\Theta \to -2$, \textbf{Ric}$[\widehat{\bm g}]$ also has an extra term $2\, {}^3 R$ of the $\bm u$-foliation. We highlight similar foliation dependent terms in the full Riemann tensor.
\\
\\
\noindent We present some explicit examples for FLRW spacetimes in standard foliation and spherically symmetric spacetimes in the Painleve-Gullstrand foliation. We briefly discuss implications of the results for Euclidean quantum gravity and quantum cosmology.
\end{abstract}

\maketitle
\vskip 0.5 in
\noindent
\maketitle
\section{Introduction} \label{sec:intro} 
Euclidean manifolds play an important role in the path-integral approach to quantum gravity, which forms the basis of the Euclidean quantum gravity program \cite{book-gh}. 
%
%
In the standard approach, the Euclidean metrics that are considered are obtained from the Lorentzian one by the process of Wick rotation,  $t \to - it$. However, while this works well in flat/static spacetime, it is ambiguous when arbitrary coordinates are used, and even more so in a general curved spacetime. Moreover, the physical significance of Wick rotation used to analytically continue the action is not entirely clear -- the gravitational case presents issues which do not arise when a similar procedure is applied to other gauge theories. In addition, there is no straightforward way of doing the analytic continuation while still keeping the metric $\bm g$ real when the metric components depend on time explicitly, and/or the metric contains cross terms such as ${\bm \DM} t \otimes {\bm \DM} x^a$. Of all the issues that Euclidean quantum gravity faces, these are the most fundamental, and they are basically related to the problem of performing Wick rotation in a generic curved spacetime. 

{\it Observer dependence}:

But the problem, in a sense, indicates a possible solution as well. The key idea of Euclidean QG of Wick rotating the time coordinate $t$ could itself suggest the resolution. Perhaps one should focus not on $t$ itself but rather on a set of timelike curves along which $t$ measures the proper time. The set of curves which can correspond to a sensible definition of $t$ would define a foliation $\Sigma$, characterised by the tangent $\bm u$ to the curves. And indeed, as pointed out by Hawking \& Ellis \cite{book-he}, given a non-vanishing timelike vector field $\bm u$, one can construct a Euclidean metric \textcolor{black}{$\bm g_E^{-1} = \bm g^{-1} + 2 \bm u \otimes \bm u$}. Condition under which a given Euclidean manifold admits a Lorentzian metric  (or the converse, which is more relevant for our case) therefore reduces to that of existence of a smooth, nowhere vanishing vector field $\bm u$. Such a vector field always exists for non-compact manifolds, while compact manifolds admit one iff their Euler number is zero.

Let us highlight two key advantages of this simple modification. First, it helps define a Euclidean geometry corresponding to a given Lorentzian geometry, at least in regions of the Lorentzian manifold which admit a non-vanishing timelike vector field. There is no ambiguity about this procedure, in particular, the metric components remain real. Of course, the catch is that one now has to choose a particular vector field $\bm u$, and the Euclidean geometry obtained would depend on this choice. This, however, turns out to be a feature rather than a problem, and brings us to the second advantage. The dependence on $\bm u$ in fact lets us introduce the notion of observer dependence, a notion that has repeatedly asserted its significance in the study of thermal effects associated with spacetime horizons. Indeed, since quantum description is inherently observer dependent, it is extremely plausible that any fundamental framework of quantum gravity would need to have this observer dependence built into it structure (implicitly or explicitly) if it has to correctly reproduce the results from semiclassical gravity \cite{book-birrell} in appropriate limit. Of course, it is also possible that such a limit is non-trivial and the semiclassical results simply disappear in the final theory of quantum gravity, rendered merely as artefacts of certain approximations made. However, most of these results really only require basic principles of free field theory and general relativity that have been tested to a great accuracy, and hence it is more likely that these results will hold in the full theory of quantum gravity as well. In fact, it was pointed out long back by Calzetta and Kandus that quantum cosmology inherits the observer dependence of vacuum in quantum field theory, through the choice of Wick rotation \cite{calzetta-kandus}.

Therefore, the dependence of the Euclidean metrics associated with a Lorentzian spacetime $\l( \mathcal{M}, \bm g \r)$ on a vector field $\bm u$ serves to introduce the important notion of {\it observer dependence}, a notion that is inherent to quantum field theory and hence to the full framework of quantum gravity as well. 
\textcolor{black}{
The relevant field space for quantum dynamics now becomes $\mathcal{F}_{\bm u} = \l\{ \l( \mathcal{M}, \bm g, \bm u \r) {\large |} \bm g(\bm u, \bm u)=-1 \r\}$.
Of course, this raises important conceptual points. For instance, one may ask: Is the spacetime ``really" Euclidean rather than Lorentzian at small scales? Though a subtle point \cite{sorkin}, as far as the key aim of this paper is concerned it is somewhat secondary to the more relevant aspect: the form of the action in the domain where the metric $\widehat{\bm g}$ has Euclidean signature -- the domain usually attributed to quantum phenomenon such as tunnelling. However, the Lorentzian metric still plays the key role, and I will express the Euclidean action in terms of Lorentzian quantities. Another pertinent question is: Are there observables insensitive to the choice of $\bm u$? This can be addressed by careful inspection of the curvature tensor for $\ge$ (given below), but more work is needed to give a satisfactory answer \cite{observer-space, samuel}. 
}

\section{Metrics describing Euclidean to Lorentzian transition} \label{sec:ge}
However, $\bm g_E$ itself is not of much use. The spacetime at large scales is not Euclidean but Lorentzian, so what we really need is a family of metrics that interpolate between a Euclidean geometry at small scales and a Lorentzian one at larges scales. What do we mean by scale here? Since $\bm g_E$ depends on $\bm u$, the only sensible operational notions of small and large scales is through the affine parameter $\lambda$ along $\bm u$ that satisfies
\begin{eqnarray}
\bm u \cdot \nabla \lambda = 1
\end{eqnarray}
\textcolor{black}{Here, $\lambda$ is essentially the proper time along $\bm u$, but instead of conventional Wick rotation,} we now characterise the Euclidean to Lorentzian transition by considering a class of metrics
\begin{eqnarray}
\bm \ge^{-1} = \bm g^{-1} - \Theta \, \bm u \otimes \bm u
\end{eqnarray}
where $\Theta(\lambda)$ is a transition function who precise details we will not need here. The metric tensor itself is given by
\begin{eqnarray}
\bm \ge = \bm g + \frac{\Theta}{1+\Theta} \, \bm t \otimes \bm t
\end{eqnarray}
where $\bm t = \bm g \l(\bm u, \cdot \r)$ is the one form associated with $\bm u$. 
\footnote{Feynman propagator \textcolor{black}{for a free scalar field} with a similar one-parameter class of metrics was discussed by Candelas \& Raine, and Visser \cite{candelas-raine-visser}. In their work, $\Theta$ was assumed to be a constant parameter, while we require it to be a function.} 

As expected, $\bm \ge$ is degenerate and is singular at $\Theta=-1$; as we shall see, the hypersurface $\Sigma_0$ corresponding to this is the boundary between the Euclidean and the Lorentzian regimes described by $\bm \ge$.
	\begin{figure}[!htb]
	\begin{center}
	\scalebox{0.35}{\includegraphics{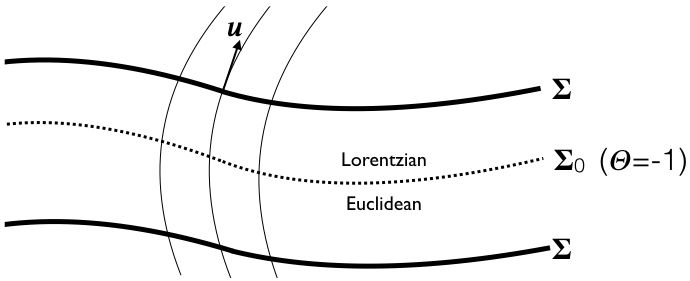}}
	\end{center}
	\caption{Euclidean to Lorentzian transition described by $\bm \ge$.}
	\label{fig:transition}
	\end{figure}
There is no reason for the metrics described by $\bm \ge$ to be equivalent to either $\bm g$ or to the ones obtained from it via Wick rotation $t \rightarrow -it$. This will be immediately evident form the curvature tensors associated with $\bm \ge$ that we discuss in Section \ref{sec:curvature}. Before proceeding to do this, we first discuss 
an example which is simple to visualize, and might also be important for understanding the role, in quantum cosmology, of spaces with $\Lambda<0$ in yielding a universe with $\Lambda>0$.
\subsection{Maximally symmetric space(time)s} \label{sec:dS-EAdS}
Consider the de Sitter spacetime ($\Lambda>0$) with flat slicing, given by $\DM s^2 = - \DM t^2 + \exp{(2t/\ell)} \delta_{\mu \nu} \DM x^\mu \DM x^\nu$, with the geodesic timelike vector field $\bm u = \partial_t$. In this case, it is trivial to see that the line element corresponding to $\bm \ge$ is given by $\widehat{\DM s}^2 = - (1+\Theta)^{-1} \DM t^2 + \exp{(2t/\ell)} \delta_{\mu \nu} \DM x^\mu \DM x^\nu$. This is clearly different from what one would obtain from Wick rotation (which in this case would yield a complex metric). For this particular case, the geometric nature of $\bm \ge$ is best described by considering the embedding of $\bm \ge$ in $D=5$ Minkowski spacetime for $\Theta=0$ and $\Theta=-1$. These describe, respectively, the $D=4$ de Sitter spacetime ($\DM{S}_4$) and $D=4$ Euclidean anti-de Sitter space (the hyperbolic space $\mathds{H}^4$), which are hyperboloids in the ambient space given by 
	\begin{figure}[!htb]
	\begin{center}
	\scalebox{0.35}{\includegraphics{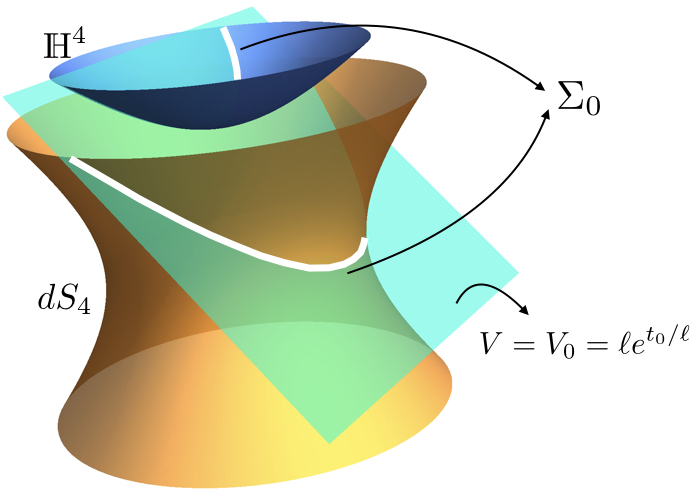}}
	\end{center}
	\caption{Embedding space description of transition between Euclidean anti-de Sitter and de Sitter described by $\bm \ge$. The thick white lines describe the 
	transition hypersurface $\Sigma_0$.}
	\label{fig:transition}
	\end{figure}
\begin{eqnarray}
\epsilon U V + \sum \limits_{1}^{3} {X^\mu}^2 = \ell^2
\end{eqnarray}
with $\epsilon=-1$ for $\DM{S}_4$ and $\epsilon=+1$ for $\mathds{H}^4$. $U=T-Z$, $V=T+Z$ are the standard null coordinates on Minkowski spacetime in standard flat coordinates $(T, Z, X^1, X^2, X^3)$. The embedding is given by 
\begin{eqnarray}
V &=& \ell e^{t/\ell}
\nn \\
U &=& \epsilon \ell e^{-t/\ell} + (1/\ell) e^{t/\ell} \sum \limits_{1}^{3} {x^\mu}^2
\nn \\
X^\mu &=& \l(V/\ell\r) x^\mu
\end{eqnarray}
The induced metric is then given by 
$$\DM s^2 = \epsilon \DM t^2 + \exp{(2t/\ell)} \delta_{\mu \nu} \DM x^\mu \DM x^\nu$$
which is equivalent to $\bm \ge$ with $1/(1+\Theta) = - \epsilon$.

The $t=$ constant slices correspond to the null plane $V=$constant of the Minkowski space, and the transition hypersurface $\Sigma_0$ in this case is connected by a null plane in the ambient space. As expected, the metric is therefore generate on this hypersurface.

In this particular example, our proposed transformation maps a maximally symmetric spacetime (dS) to another maximally symmetric spacetime (EAdS) in 
the Euclidean regime ($\Theta \to -2$), with curvature changing sign \cite{hhh-qc-ads}. However, this only happens in the $k=0$ (flat) foliation of dS. We will revisit the $k \neq 0$ case below after deriving the full expression for the Riemann tensor of $\bm \ge$.
\\
\section{The geometry described by $\ge$} \label{sec:curvature} 
It is straightforward, although lengthy, to write down the curvature tensor associated with $\bm \ge$ in terms of the quantities associated with $\bm g$ and those describing the intrinsic and extrinsic geometry of $\bm u$ foliation, on which the induced metric is (the projection of) $h^a_{\phantom{a} b}=\delta^a_{\phantom{a} b} + u^a t_{b}$. 
Some simplications happen upon using the Gauss-Codazzi and Gauss-Weingarten equations 
, and the final form turns out to be
\\
\begin{widetext}
\begin{eqnarray}
\widehat{R}^{ab}_{\phantom{ab}cd} &=& {R}^{ab}_{\phantom{ab}cd} \;+\; 2 \Theta \; \Biggl[ t_m R^{m[a}_{\phantom{m[a}cd} \; u^{b]} + K^{[a}_{\phantom{[a}[c} K^{b]}_{\phantom{b]}d]} \Biggl] \;+\; 
2 \l( \frac{\DM \Theta}{\DM \lambda} \r) \; u^{[a} K^{b]}_{\phantom{b]}[c} \; t_{d]}
\label{eq:riemann-gen}
\end{eqnarray}
from which the following can be derived in a straightforward manner
\begin{eqnarray}
\widehat{R}^{a}_{\phantom{a}b} &=& \l(1 + \Theta \r) {R}^{a}_{\phantom{a}b} - \Theta \Biggl[ {}^{(3)}R^a_{\phantom{a}b} - t_b \mathcal{C}^a \Biggl] \;+\; 
\frac{1}{2} \l( \frac{\DM \Theta}{\DM \lambda} \r) \; \Biggl[ \pi^a_{\phantom{a}b} + K \delta^a_b \Biggl]
\\
\widehat{G}^{a}_{\phantom{a}b} &=& \l(1 + \Theta \r) {G}^{a}_{\phantom{a}b} - \Theta \Biggl[ {}^{(3)}G^a_{\phantom{a}b} + \frac{1}{2} \; {}^{(3)}R \; u^a t_b - t_b \mathcal{C}^a \Biggl] \;+\; 
\frac{1}{2} \l( \frac{\DM \Theta}{\DM \lambda} \r) \; \pi^a_{\phantom{a}b}
\label{eq:ricci-einstein-gen}
\end{eqnarray}
\end{widetext}
where $\mathcal{C}^a = R^m_{\phantom{m}n} q^n h^a_m = D_m K^{ma} - D^a K$ and $\pi^a_{\phantom{a}b} = K^a_b - K h^a_b$. 
The Ricci scalar is given by \cite{dk-grg}
\begin{eqnarray}
\widehat{R} &=&  \l(1 + \Theta \r) R - \Theta \; {}^{(3)}R +  \l( \frac{\DM \Theta}{\DM \lambda} \r) K
\label{eq:ricci-scalar-gen}
\end{eqnarray}

Before we discuss the implications of the above expressions, we highlight their behaviour on the hypersurface defined by $\Theta=-1$. The metric expectedly becomes degenerate here. However, the following limit holds:
\begin{eqnarray}
\lim \limits_{\Theta \rightarrow -1} \widehat{R}^{ab}_{\phantom{ab}cd} \, e^{(\mu)}_a e^{(\nu)}_b e^c_{(\rho)} e^d_{(\sigma)} &=& {}^{(3)}{R}^{\mu \nu}_{\phantom{\mu \nu}\rho \sigma}
\end{eqnarray}
yielding similar limits for all the other tensors; in particular, $\lim \limits_{\Theta \rightarrow -1} \widehat{R} = {}^{(3)}R$.

\textcolor{black}{Let me briefly comment on the above expressions vis-a-vis the 3+1 split $\bm g = - N^2 \bm \DM t \otimes \bm \DM t + h_{\mu \nu} (\bm \DM x^\mu + N^\mu \bm \DM t) \otimes (\bm \DM x^\nu + N^\nu \bm \DM t)$. In terms of $\bm h = \bm g + \bm t \otimes \bm t$, we have $\ge = \bm h - (1+\Theta)^{-1} \, \bm t \otimes \bm t$, which makes evident the degenerate nature of $\ge$ at $\Theta=-1$. Further, the usual Gauss-Codazzi decomposition involves terms quadratic in $K_{ab}$, while the $\Theta$ dependence here leads also to terms linear in $K_{ab}$.
}
\section{A step profile for $\Theta(\lambda)$} \label{sec:step}
The expressions given above for the curvature tensor and its concomitants display some remarkable characteristics, manifest in particular from the terms related to intrinsic geometry of $\Sigma$ that couple to $\Theta$ and $\dot{\Theta} = \DM \Theta/\DM \lambda$. To investigate further, we must discuss the profile of the transition function $\Theta(\lambda)$. Of course, we can not be too restrictive about the form of this function {\it a priori}. The only requirement we can put on it is that, being dimensionless, it depends only on the ratio $x = (\lambda/\lp)$, where $\lp$ is some fundamental length scale that characterises the transition. There might be another length scale $w_0$ that characterises the ``width" of the transition, we ignore it here. Further, to effect the transition from Euclidean to Lorentzian, it must satisfy the following limits

\begin{eqnarray}
\lim \limits_{x \rightarrow 0} \Theta(x) &=& - 2
\nn \\
\lim \limits_{x \rightarrow \infty} \Theta(x) &=& \phantom{-} 0
\end{eqnarray}

\begin{figure}[!htb]
	\begin{center}
	\scalebox{0.5}{\includegraphics{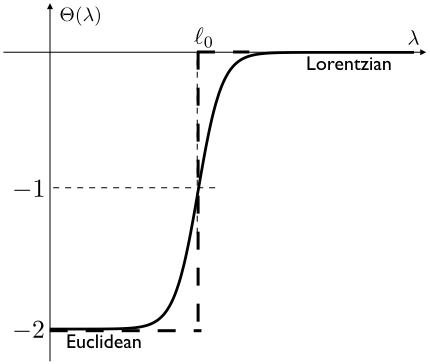}}
	\end{center}
	\caption{A typical profile for the transition function $\Theta(\lambda)$. The dashed curve is an approximation by a step profile, the case discussed in detail in this paper.}
	\label{fig:theta}
	\end{figure}

A typical profile of this form is depicted in Fig. \ref{fig:theta}.

As an aside, let us point out that the above profile for $\Theta(x)$ is reminiscent of soliton solutions in classical field theory, and, in fact, if one can find such soliton like solutions in a curved background, these could trigger the transition from Euclidean to Lorentzian spacetimes discussed in this paper.

In what follows, we shall focus on the sharp transition and describe the function in terms of the Heaviside step function $\theta(x)$ as
\begin{eqnarray}
\Theta_0(\lambda) &=& 2 \theta(\lambda-\lp) - 2
\end{eqnarray}
which is essentially the $w_0 \rightarrow 0$ limit of the form discussed above. We then have $\dot{\Theta}_0(\lambda) = 2 \delta(\lambda-\lp)$, and the various expressions for curvature tensors simplify considerably. For simplicity, we shall write $\delta(\lambda-\lp)$ as $\delta_{\Sigma_0}$ in the expressions below.
\begin{eqnarray}
\widehat{R}^{ab}_{\phantom{ab}cd} &=& {R}^{ab}_{\phantom{ab}cd} \;+\; 2 \Theta_0 \; \Biggl[ t_m R^{m[a}_{\phantom{m[a}cd} \; u^{b]} + K^{[a}_{\phantom{[a}[c} K^{b]}_{\phantom{b]}d]} \Biggl] 
\nn \\
&& \phantom{-------------} \;+\; 
4 \delta_{\Sigma_0} \; u^{[a} K^{b]}_{\phantom{b]}[c} \; t_{d]}
\nn \\
\widehat{R}^{a}_{\phantom{a}b} &=& \l(1 + \Theta_0 \r) {R}^{a}_{\phantom{a}b} - \Theta_0 \Biggl[ {}^{(3)}R^a_{\phantom{a}b} - t_b \mathcal{C}^a \Biggl] 
\nn \\
&& \phantom{-------------} \;+\; 
\delta_{\Sigma_0} \; \Biggl[ \pi^a_{\phantom{a}b} + K \delta^a_b \Biggl]
\nn \\
\end{eqnarray}
Of particular interest are the forms of the Ricci scalar 
\begin{eqnarray}
\widehat{R} &=&  \l(1 + \Theta_0 \r) R - \Theta_0 \; {}^{(3)}R +  \delta_{\Sigma_0} \; (2 K)
\end{eqnarray}
and the Einstein tensor 
\begin{eqnarray}
\widehat{G}^{a}_{\phantom{a}b} &=& \l(1 + \Theta_0 \r) {G}^{a}_{\phantom{a}b} - \Theta_0 \Biggl[ {}^{(3)}G^a_{\phantom{a}b} + \frac{1}{2} \; {}^{(3)}R \; u^a t_b - t_b \mathcal{C}^a \Biggl] 
\nn \\
&& \phantom{---------} \;+\; 
\delta_{\Sigma_0} \; \l( K^a_{\phantom{a}b} - K h^a_{b} \r)
\nn \\
\end{eqnarray}
which have the limiting forms

{\it Lorentzian regime} ($\Theta_0=0$):
\begin{eqnarray}
\widehat{R} &=& R + (2 K) \delta_{\Sigma_0} 
\nn \\
\widehat{G}^{a}_{\phantom{a}b} &=& {G}^{a}_{\phantom{a}b} + 
\l( K^a_{\phantom{a}b} - K h^a_{b} \r) \delta_{\Sigma_0}
\nn \\
\end{eqnarray}
{\it Euclidean regime} ($\Theta_0=-2$):
\begin{eqnarray}
\widehat{R} &=& - R + 2 \; {}^{(3)}R + (2 K) \delta_{\Sigma_0} 
\nn \\
\widehat{G}^{a}_{\phantom{a}b} &=& - {G}^{a}_{\phantom{a}b} + 2 \Biggl[ {}^{(3)}G^a_{\phantom{a}b} + \frac{1}{2} \; {}^{(3)}R \; u^a t_b - t_b \mathcal{C}^a \Biggl] 
\nn \\
&& \phantom{---------} \;+\; 
\l( K^a_{\phantom{a}b} - K h^a_{b} \r) \delta_{\Sigma_0}
\nn \\
\label{eq:ricci-einstein-euclidean}
\end{eqnarray}

The above analysis highlights several non-trivial points concerning gravitational dynamics associated with Euclidean metrics that might arise naturally while studying quantum aspects of gravity and the associated small scale structure of spacetime. We summarise the key ones below:
	\begin{itemize}
	\item The terms involving $\dot{\Theta}$ in curvature tensors have a natural interpretation in terms of known quantities; for 
	the Ricci scalar depends, this term precisely yields the correct GHY boundary term, while for the Einstein tensor, it gives the 
	canonical momentum associated with the $\bm u$ foliation.
	\item In the Lorentzian regime, the Ricci scalar for $\bm \ge$ is equivalent to the {\it full} Einstein-Hilbert lagrangian with the 
	correct surface term.
	\item More importantly, in the Euclidean regime, the Ricci scalar is not simply $(-R)$, even ignoring for the moment the 
	boundary term. It involves an addition term involving ${}^{(3)}R$. Similarly, the Einstein tensor also has additional terms 
	involving quantities defined on $\Sigma$. The relevance of this becomes evident when $\bm g$ is such that 
	$R=0=G^a_{\phantom{a}b}$ (that is, any vacuum solution), and the additional terms are all that remain.
	\end{itemize}
\section{Examples} \label{sec:eg}
In this section, we apply the expressions derived above to study the Euclidean regimes of some relevant spacetimes in physically 
interesting foliations.

	\subsection{Maximally symmetric foliations of Maximally symmetric space(time)s} \label{sec:maxsymm}
Using the above expressions, it is easy to obtain results for maximally symmetric ($t=$ constant in the metric given below) foliations of maximally symmetric spacetimes $(\mathcal{M}, \bm g)$ described by 
\begin{eqnarray}
\DM s^2 = - \DM t^2 + a(t)^2 \DM \Omega_{(k)}^2 \hspace{1cm} (k=-1, 0, 1)
\end{eqnarray} 
which are of obvious relevance to cosmology. We already studied one example of this case in \ref{sec:dS-EAdS} (the case of flat foliation of de Sitter spacetime). We now present the general case. For these spacetimes, $K^a_{\phantom{a}b} = H(t) h^a_{\phantom{a}b}$, where $H(t)\dot{a}/a$, and ${R}^{ab}_{\phantom{ab}cd} = \ell^{-2} \l( \delta^a_c \delta^b_d - \delta^a_d \delta^b_c \r)$ where $\ell$ is the curvature scale of the spacetime, and $a(t)$ is a specific function depending on $k$; for example, for $k=0$, $a(t)=a_0 e^{t/\ell}$. It is then straightforward to show that

{\it Euclidean regime} ($\Theta_0=-2$):
\begin{eqnarray}
\widehat{R}^{ab}_{\phantom{ab}cd} &=& \l( 1 - 2 H^2 \ell^2 \r) {R}^{ab}_{\phantom{ab}cd} \;-\; \frac{8 (1 - H^2 \ell^2)}{\ell^2} t_{[c} \delta^{[a}_{d]} q^{b]}
\nn \\
\end{eqnarray}
It is evident that only for flat slicing ($k=0$), $H(t) =$ constant $= 1/\ell$, and the resultant spacetime is again maximally symmetric with $\widehat{R}^{ab}_{\phantom{ab}cd} = - {R}^{ab}_{\phantom{ab}cd} + $ (surface term). For the other slicings, the metric in the Euclidean regime is no longer maximally symmetric.

The structure of $\widehat{G}^{a}_{\phantom{a}b}$ is also interesting to note. Since the slicing is maximally symmetric, one has ${}^{(3)}G^a_{\phantom{a}b} = - ({1}/{6}) \; {}^{(3)}R$ and $\mathcal{C}^a=0$. Therefore, using Eq. (\ref{eq:ricci-einstein-gen}), we obtain
\begin{eqnarray}
\widehat{G}^{a}_{\phantom{a}b} &=& \l(1 + \Theta \r) {G}^{a}_{\phantom{a}b} + \Theta \l( \frac{1}{6} {}^{(3)}R \r) \Biggl[ h^a_{\phantom{a}b} - 3 u^a t_b \Biggl] 
\nn \\
&& \phantom{---------} \;+\; 
(1/2) \dot{\Theta} \; \pi^a_{\phantom{a}b}
\end{eqnarray}

	\subsection{Foliation by radial timelike geodesics in Spherically symmetric space(time)s} \label{sec:sphsymm}
Another case of interest are foliations by radially ingoing timelike geodesics in spherically symmetric spacetimes. The coordinate system best suited for this purpose is the Painleve-Gullstrand (PG) coordinate system, which has been studied extensively for a large class of spherically symmetric spacetimes. We will consider spherically symmetric spacetimes with metric $\bm g$, such that $\DM s^2 = -f(r) \DM t^2 + (1/f(r)) \DM r^2 + r^2 \DM \Omega^2$, and focus on the one-parameter class of radially ingoing timelike geodesics $\bm u$ characterized by a real number $$0<p\leq1$$ 
This parameter is related to the conserved {\it energy at infinity} by $p=1/E_{\infty}^2=1-v_{\infty}^2$. The case $p=1$, corresponding to $E_\infty=1, v_{\infty}=0$,
yields a family of geodesics which all start from rest at infinity, the case usually discussed in textbooks. The other limit, $p \to 0$, corresponds to $E_\infty \to \infty, v_{\infty}=1$, and corresponds to a family of geodesics starting from infinity with the speed of light. In this case, it can be shown that the above coordinates reduce to the Eddington-Finkelstein ingoing coordinates \cite{poisson-martel}. 

The vector field $\bm u$ is given by $\bm u = (1/f\sqrt{p}) {\bm \partial}_t - (\sqrt{1-pf}/\sqrt{p}) {\bm \partial}_r$. In PG coordinates, the metric is given by
\begin{eqnarray}
\DM s^2 = - f \DM T^2 + 2 \sqrt{1-pf} \DM T \DM r + p \DM r^2 + r^2 \DM \Omega^2
\end{eqnarray} 
where $\DM t = \DM T - f^{-1} \sqrt{1-pf} \DM r$ defines the PG time coordinate $T$.

It is evident that the $T=$ constant slices are not flat unless $p=1$. It is easy to show that for these slices, 
\begin{eqnarray}
{}^{(3)}R &=& - \frac{2}{r^2} \l( E_{\infty}^2 - 1 \r)
\nn \\
{}^{(3)}G^a_{\phantom{a}b} &=& \phantom{+} \frac{1}{r^2} \l( E_{\infty}^2 - 1 \r) \; \delta^a_r \delta^r_b
\end{eqnarray} 
Due to spherical symmetry, the extrinsic curvature corresponding to $T= $ constant slices is given by
\begin{eqnarray}
K_{ab} &=& K_1(r) \sigma_{ab} + K_2(r) n_a n_b
\nn \\
K &=& 2 K_1(r) + K_2(r)
\end{eqnarray} 
where
\begin{eqnarray}
K_1(r) &=& - \frac{1}{r^{3/2}} \sqrt{a + \l( E_{\infty}^2 - 1 \r) r}
\nn \\
K_2(r) &=& + \frac{1}{2 r^{3/2}} \frac{a}{\sqrt{a + \l( E_{\infty}^2 - 1 \r) r}}
\end{eqnarray} 
where $n_a = \sqrt{p} \partial_a r$ are normals to $r= $ constant surfaces and $\sigma_{ab}=g_{ab} + u_a u_b - n_a n_b$ is the corresponding induced metric. It is worth taking  a pause here to discuss two important limits of extrinsic curvature (we focus on $K$):
\begin{eqnarray}
\lim \limits_{E_{\infty} \to 1} K &=& - \frac{3}{2} \frac{\sqrt{a}}{r^{3/2}} \;\;\;\;\; ({\rm Painleve}\mbox{-}{\rm Gullstrand})
\nn \\
\lim \limits_{E_{\infty} \to \infty} K &=& - \frac{2 E_{\infty}}{r} \;\;\;\;\; ({\rm Eddington}\mbox{-}{\rm Finkelstein})
\end{eqnarray} 
(The second expression clearly indicates that $\lim \limits_{E_{\infty} \to \infty} \l( K/E_{\infty} \r)$ is equal to the expansion of the ingoing null congruence.)

Coming back to the discussion of curvature tensors, for vacuum solutions $G^a_{\phantom{a}b}=0=\mathcal{C}^a$, we have

{\it Lorentzian regime} ($\Theta_0=0$):
\begin{eqnarray}
\widehat{R} &=& (2 K) \delta_{\Sigma_0} 
\nn \\
\widehat{G}^{a}_{\phantom{a}b} &=& \pi^a_{\phantom{a}b}\; \delta_{\Sigma_0}
\nn \\
\end{eqnarray}
{\it Euclidean regime} ($\Theta_0=-2$):
\begin{eqnarray}
\widehat{R} &=& - \frac{4}{r^2} \l( E_{\infty}^2 - 1 \r) + (2 K) \delta_{\Sigma_0} 
\nn \\
\widehat{G}^{a}_{\phantom{a}b} &=& +  \frac{2}{r^2} \l( E_{\infty}^2 - 1 \r) \Biggl[ \delta^a_r \delta^r_b - u^a t_b \Biggl] 
\;+\; 
\pi^a_{\phantom{a}b}\; \delta_{\Sigma_0}
\nn \\
\end{eqnarray}

\section{Implications and Discussion} \label{sec:implications} 
The results presented here bring to light several mathematical aspects of Euclidean quantum gravity which have not been noticed or discussed in the literature.
\begin{enumerate}
\item The Euclidean action usually considered in the literature is based on the Ricci scalar as the lagrangian. However, as our results show, this may not be the correct choice in the Euclidean regime. In fact, in the Euclidean regime $\Theta=-2$, we obtain (ref. Eqs. (\ref{eq:ricci-einstein-euclidean})) 
\begin{eqnarray}
\widehat{R} &=& - R + 2 \; {}^{(3)}R + (2 K) \delta_{\Sigma_0} 
\nn \\
&=& + R + 2 \; \l( K_{ab}^2 - K^2 + 2 R_{ab} u^a u^b \r) + (2 K) \delta_{\Sigma_0} 
\end{eqnarray}
In particular, even if $R=0$, $\widehat{R} \neq 0$, and is given by an action involving $K_{ab}$ and $R_{ab} u^a u^b$.
This point may be important for several reasons. 

In quantum cosmology, the Hawking-Hartle prescription for the ground state wave function of the universe is through the path integral over Euclidean geometries that have $\Sigma_0$ as their only boundary. It is obvious that the additional terms above will affect the exact details of the wave function so calculated, since the on-shell value of the action itself now depends on $ {}^{(3)}R$.

\item The partition function $Z$ for quantum gravity, based on the class of space(time)s described by $\bm \ge$, can be written as the path integral
\begin{eqnarray}
Z = \int \mathcal{D} \bm g \; \mathcal{D} \bm u \; \exp{\l[ - i \int \widehat{R} \sqrt{- {\rm det}\, \bm \ge} \r]}
\end{eqnarray}
Note that, since
$$
{\rm det}\, \bm \ge = (1+\Theta)^{-1} \, {\rm det}\, \bm g
$$
and ${\rm det}\, \bm g<0$, $\sqrt{- {\rm det}\, \bm \ge}$ is imaginary for $\Theta=-2$ (in general, for $\Theta < -1$), and hence appropriate branch cut must be chosen for the square root to make the path-integral for $Z$ convergent.

We might also highlight here a point emphasized earlier by Visser (see the second reference in \cite{candelas-raine-visser}). By construction, our entire formulation is based on a Lorentzian metric $\bm g$ and a suitable timelike vector field $\bm u$. The Euclidean regime of resultant metrics $\bm \ge$ therefore will comprise of only those manifolds that are compatible with existence of a Lorentzian structure. Therefore, it makes sense that the functional integral (in the Euclidean regime) is taken not over all Euclidean manifolds, but only those compatible with the existence of a Lorentzian structure.

\item The approach suggested here might also be relevant for the discussion of small scale structure of spacetime. In particular, one may take motivation from Euclidean quantum gravity applied to early universe, and ask if the same ideas can be applied to (i) black hole singularities in particular (in this context, the results for spherically symmetric spacetimes discussed in Sec.\,\ref{sec:sphsymm} might be of use), and (ii) spacetime in general, when probed at very small scales. For example, given any spacetime event $p$, one might consider the set of points $q \in I^+(p)$ that lie at the constant geodesic distance from $p$. Such a foliation has recently been studied and used in the investigation of the small scale structure of spacetime in presence of a minimal length \cite{dk-minimal-length}, and it is worth exploring the structure of $\bm \ge$ for the same. 
\end{enumerate}

%

\end{document}